\documentclass[aps, pra, twocolumn, notitlepage,superscriptaddress,nofootinbib]{revtex4-2}

\usepackage{amssymb,latexsym,amsmath}
\usepackage{bm}
\usepackage{amssymb,amsmath}
\usepackage{graphicx,xcolor}
\usepackage[titletoc, title]{appendix}
\usepackage[caption=false,lofdepth,lotdepth]{subfig}
\usepackage{hyperref}   
\hypersetup{%
	pdfpagemode=FullScreen,
	pdfstartpage=1,
	pdfmenubar=true,
	pdftoolbar=true,
	colorlinks = true,
	linkcolor=blue,
	citecolor=blue,
	urlcolor=blue,
	bookmarksopen=false
}

\usepackage{soul}
\usepackage[normalem]{ulem}

\begin{document}

\title{Mass-driven vortex collisions in flat superfluids}
\author{Andrea Richaud}
\email{arichaud@sissa.it}
\affiliation{Scuola Internazionale Superiore di Studi Avanzati (SISSA), Via Bonomea 265, I-34136, Trieste, Italy}

\author{Giacomo Lamporesi}
\affiliation{INO-CNR BEC Center and Dipartimento di Fisica, Universit\`a di Trento, 38123 Povo, Italy}

\author{Massimo Capone}
\affiliation{Scuola Internazionale Superiore di Studi Avanzati (SISSA), Via Bonomea 265, I-34136, Trieste, Italy}
\affiliation{CNR-IOM Democritos, Via Bonomea 265, I-34136 Trieste, Italy}

\author{Alessio Recati}
\affiliation{INO-CNR BEC Center and Dipartimento di Fisica, Universit\`a di Trento, 38123 Povo, Italy}

\date{\today}

\begin{abstract}
Quantum vortices are often endowed with an effective inertial mass, due, for example, to massive particles in their cores. Such ``massive vortices" display new phenomena beyond the standard picture of superfluid vortex dynamics, where the mass is neglected. In this work, we demonstrate that massive vortices are allowed to collide, as opposed to their massless counterparts.  We propose a scheme to generate controllable, repeatable, deterministic
collisional events in pairs of quantum vortices.
We demonstrate two mass-driven fundamental processes: (i) the annihilation of two counter-rotating vortices and (ii) the merging of two co-rotating vortices, thus pointing out new mechanisms supporting incompressible-to-compressible kinetic energy conversion, as well as doubly-quantized vortices stabilization in flat superfluids.

\end{abstract}

\maketitle

Quantum vortices are topological excitations characterized by the quantized circulation of the velocity field. Their appearance in a rotating system represents the ``smoking gun" of superfluidity, since they are the most direct consequence of the existence of a macroscopic wavefunction associated to the superfluid \cite{Onsager1949}. They are observed in contexts as diverse as liquid Helium \cite{Donnelly,Volovik1988}, ultra-cold gases \cite{Matthews1999,Madison2000,Zwierlein2005},  superconductors \cite{Blatter1994} quantum fluids of lights in non-linear optical systems \cite{carusotto2013}. Their existence has been also speculated in neutron matter \cite{neutronstars2001,neutronstars2021}. The vortex motion is responsible for the onset of dissipative phenomena \cite{Kwon2021,Blatter1994}, while the unbinding of vortex-antivortex pairs constitutes the ultimate mechanism destroying the quasi-long-range coherence in a two-dimensional (2D) system \cite{Kosterlitz1974,Hadzibabic2006}.    

Vortices are often pictured as funnel-like holes, where the superfluid density goes to zero and around which the quantum fluid exhibits a swirling flow. However, in many physical systems, vortex cores are not simply empty, but turn out to be filled by other particles, following different mechanisms.
Such filling particles can be thermal atoms which coexist with the superfluid fraction \cite{Coddington2004}, quasiparticle bound states \cite{Kwon2021}, atoms deliberately introduced to better trace vortex trajectories \cite{Bewley2006,Griffin2020}, or also atoms of a different quantum fluid \cite{Anderson2000,Law2010,Gallemi2018,Richaud2020,Richaud2021,Ruban2021,Zhu2022}. Irrespective of their microscopic origin, atoms trapped within vortex cores have a deep influence on the vortex dynamics.

In 2D systems, vortices can be seen for many purposes as point-like objects, whose dynamics is described by an effective Lagrangian. The number $q$ characterising the quantized circulation is refereed to as the ``charge" of the vortex. If the core is empty, the dynamics does not contain any inertial term, the equations of motion are of first order and dominated by the so-called Magnus force.
The presence of a dressed core, however, makes the vortex massive and calls for the inclusion of inertial terms.

The need for the inclusion of a mass term in the dynamics of vortex-like structures (skyrmions) was discussed in the context of magnets more than 30 years ago \cite{Ivanov1989,Ivanov2005} (see Ref.~\cite{Wu2022} and references therein, for a recent discussion). Although the physical origin of the mass can be rather different from what discussed here, its effect on the skyrmion dynamics is the same.
Around 10 years ago, an inertial term was discussed by Turner \cite{Turner2009}, studying polar-core vortices in the easy-plane phase of spin-1 condensates. A Lagrangian formulation and the  comparison with numerical simulations of the full Gross-Pitaevskii equation (GPE) describing the gas can be found in Ref.  \cite{Williamson2016}.
Recently it was shown \cite{Richaud2020,Richaud2021} that the inertial term in the particle-like description of vortices arises naturally in highly unbalanced Bose-Bose mixtures, where the majority component hosts a vortex, whose core is filled by the minority one. This provided a clear interpretation of the rise of the vortex mass as due to the dynamics of the minority component.

In this work, we show that the presence of massive particles trapped within vortex cores opens the door to a new class of phenomena, when more than a single vortex is considered. While massless vortices can only interact at distance, without getting too close one to each other \cite{Neely2010,
Seo2017}, we find that massive vortices can actually collide and merge or annihilate, depending on their charge. We focus on two fundamental classes of two-vortex collisions: the collision of vortices with opposite or equal circulation. In the former case, the collision results in their mutual annihilation and in the consequent emission of sound waves. In the latter case, the collision of two singly-quantized ($q_1=q_2=\pm 1$) vortices results in the formation of a doubly-quantized vortex $q=\pm 2$, a seemingly unstable state \cite{Garcia1999,Butts1999,Castin1999,Shin2004}, which is instead stabilized by the presence of core mass, which, in turn, plays the role of an effective pinning potential \cite{Simula2002}. 
We focus on the specific case of vortices in the majority component of a strongly unbalanced quasi-2D BEC mixture. 
We propose a realistic experimental protocol, where two-vortex collisions can be observed in a controlled fashion thanks to a component-selective potential.  

We consider an atomic Bose-Bose mixture formed by $N_a$ atoms of species $a$ and $N_b$ atoms of species $b$ whose atomic masses are $m_a$ and $m_b$, respectively. The mixture is strongly confined in the $z$-direction by a harmonic potential characterized by a harmonic oscillator length $d_z$, such that the atomic motion along $z$ is frozen. At zero temperature, the gas realizes a binary BEC, characterized by the order parameters $\psi_a$ and $\psi_b$. Their dynamics obeys the coupled GPEs

\begin{equation}
\label{eq:H_GPE}
i\hbar\frac{\partial \psi_i}{\partial t} =\! \!\left(\!-\frac{\hbar^2\nabla^2}{2m_i} + V_{\rm tr}^i 
+ \sum_{j=a,b}g_{ij}N_j|\psi_j|^2\!\right)\psi_i, \; i=a,b
\end{equation}
where the quasi-2D couplings 
$g_{ij}=\sqrt{2\pi}\hbar^2 a_{ij}/(m_{ij} d_z)$ are given by intra- and inter-species s-wave scattering lengths $a_{ij}$, and effective masses $m_{ij}=1/(m_{i}^{-1}+m_{j}^{-1})$ \cite{Hadzibabic2011}.  

A particular topological solution of the coupled GPEs is a vortex in one component, whose core is filled by the other component, which has a trivial phase profile. When the system is $SU(2)$ symmetric (or close),
such solutions are named magnetic vortices since the total density is almost unaffected (see Ref.~\cite{Gallemi2018} and reference therein).
As long as the forces applied to the two components do not significantly alter the composite object, its dynamics -- as for the standard vortex in a superfluid -- can be described in terms of a Lagrangian for a point-like particle. However, a remarkable feature appears due to the filled vortex core:  the dynamics is massive, i.e., the time evolution of vortex coordinates $(x_j,\,y_j)=:{\bm r}_j$ is governed by second-order differential equations. The introduction of inertial effects in superfluid vortex dynamics or, in other words, the fact that vortices' velocities $\{\dot{\bm r}_j\}$ are independent of vortices' positions $\{{\bm r}_j\}$, opened the door to new dynamical regimes \cite{Richaud2020,Richaud2021,PRA_Griffin}. In the context of magnetic skyrmions, it has already been possible to prove that it is indeed necessary to add an effective mass to the standard Thiele  model \cite{Thiele1973},  which is inherently non inertial, in order to capture the observed trajectories \cite{Makhfudz2012,Buttner2015}.
Interestingly, in the context of half-quantum vortices in a two-component BEC, approaching trajectories of vortex dipoles have been observed by solving the GPEs \cite{Kasamatsu2016}. Since the authors used the standard point-like model in their analysis, they had to conclude by conjecturing on the possible presence of an inertial mass.

In the present work, we study the effect of the vortex mass on the vortex collisional dynamics. To ensure the vortex stability during their dynamics, we consider an immiscible mixture, corresponding to $g_{ab}^2>g_{aa}g_{bb}$ (see discussion on vortex stability in Supplemental Material). Moreover, we consider $N_b\ll N_a$, such that the vortex core is small enough and the use of an effective point-like particle description is justified. 
While in the above-mentioned works the cores play a passive role, since they merely constitute a sort of ``heavy burden" affecting the standard superfluid vortex dynamics,
here we flip the perspective and use the massive cores to actively drive vortex collisions in a controlled fashion. These collisions may indeed happen in a many-massive-vortex system,
but determining their precise location and timing is difficult and strictly dependent on the detailed initial state of the system. On top of this, the complexity of the dynamics, as well as the chaotic character of vortex trajectories, hinders the possibility to attempt any comparison between theory and experiments. 

To circumvent these difficulties, we focus on the simplest possible systems where collisions are possible: a pair of vortices with the same or opposite charge. 
This choice allows us to perform a theoretical investigation of vortex collisions, by using an experimentally accessible protocol based on a component-selective trapping potential.
We consider that the mixture is confined in a box potential of radius $R$, and that the minority component feels also a harmonic potential $V_{\mathrm{tr}}^b=m_b\omega_b^2 r^2 /2$.

\begin{figure}[b!]
    \centering
    \includegraphics[width=1\linewidth]{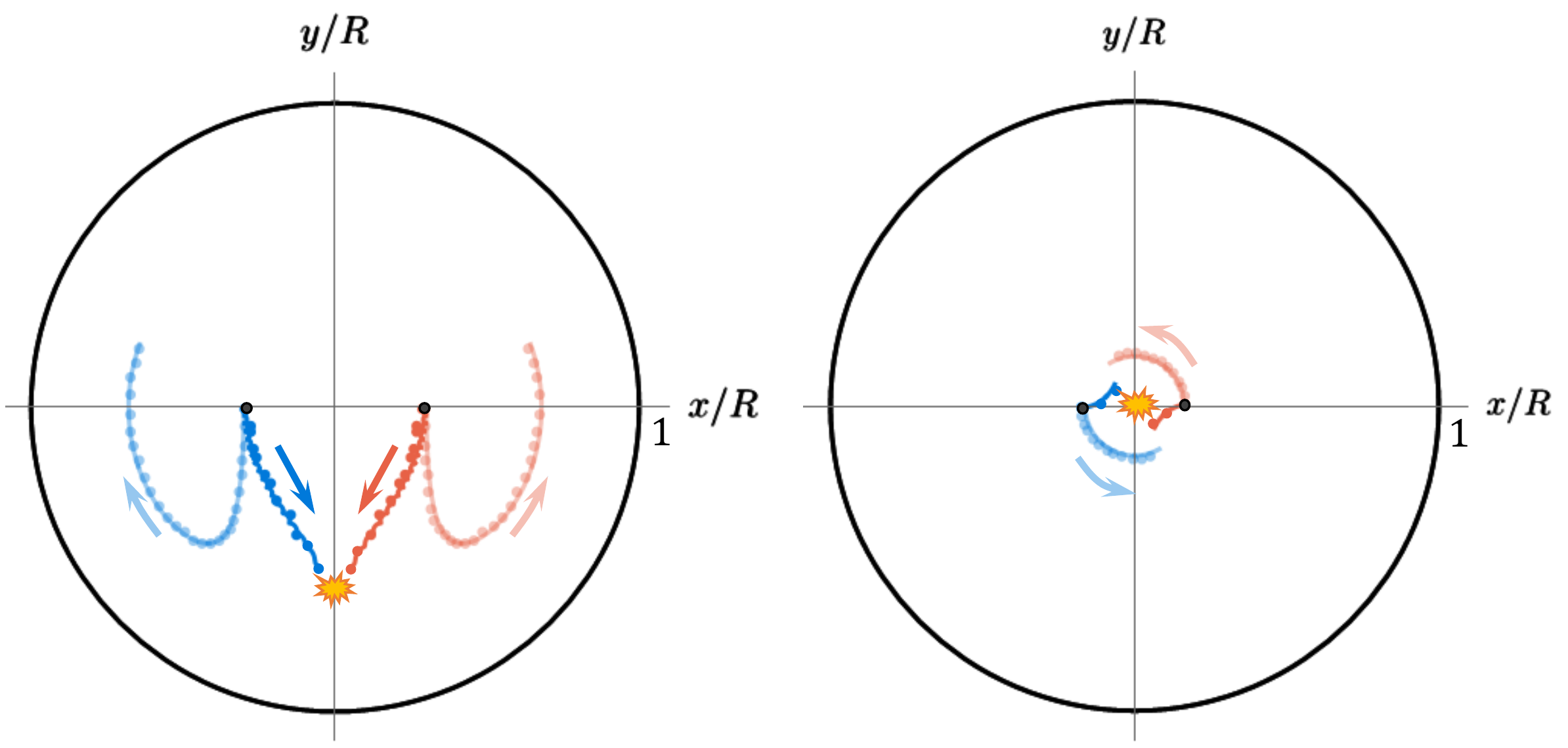}
    \caption{Comparison between the trajectories predicted by the massive point vortex model (\ref{eq:Motion_equation}) (solid lines) and the results extracted from simulations of time-dependent GPEs (dots) both in the presence (dark colors) and in the absence (light colors) of core mass. Massive cores can lead to collisions (yellow stars). Black dots represent starting positions. Left (right) panel corresponds to counter- (co-) rotating vortices. The assumed microscopic parameters are listed in the captions of Fig.~\ref{fig:Counter_rotating_collisions} and \ref{fig:Co_rotating_collisions}.}
    \label{fig:Collision_sketch}
\end{figure}

Following Ref.~\cite{Richaud2021},  one can derive the effective Euler-Lagrange equations of motion for two massive vortices subject to $V_{\mathrm{tr}}^b$ (see Supplemental Material)
$$
  M_j\ddot{\bm r}_j = k_j \rho_a \hat{z} \times \dot{\bm{r}}_j+\rho_a\frac{k_j}{2\pi}\left[k_i\frac{\bm{r}_j-\bm{r}_i}{|\bm{r}_j-\bm{r}_i|^2}+k_j\frac{\bm{r}_j}{R^2-r_j^2} \right.
$$
\begin{equation}  
\left.  +k_i \frac{R^2\bm{r}_i-r_i^2\bm{r}_j}{R^4-2R^2\bm{r}_i\bm{r}_j+r_i^2r_j^2}\right]-M_j\omega_b^2 \bm{r}_j.
\label{eq:Motion_equation}
\end{equation}
where $\rho_a=m_a n_a = m_aN_a/(\pi R^2)$ is the planar mass density of component-$a$ atoms, $k_j=q_j h/m_a$ ($q_j \in \mathbb{Z}$) is the strength of the $j$-th vortex, while $M_j=N_b m_b /2$ is the mass of component-$b$ cores (we assume, without loss of generality, that the total mass $M_b=m_bN_b$ is equally subdivided between the two vortices). The two terms $\propto \rho_a$ are well known and describe the contribution of the Magnus force, the intervortex interaction and the presence of a hard-wall confining potential (resulting in an image vortex) \cite{Kim2004}. The term $M_j\ddot{\bm r}_j$ represents the Newton-like acceleration term ensuing from the presence of massive cores \cite{Richaud2020,Richaud2021}, while the last term is due to their being subject to a harmonic confinement. The adoption of a (massive)-point-vortex model is justified for vortex cores much smaller than $R$. In our case, the size of the vortex core is of the order of the healing length $\xi_a=\hbar/\sqrt{2m_a g_{aa} n_a}$ of component $a$ and $\xi_a\ll R$ is a rather standard condition for atomic BECs.  

\begin{figure}[t!]
    \centering
    \includegraphics[width=1\linewidth]{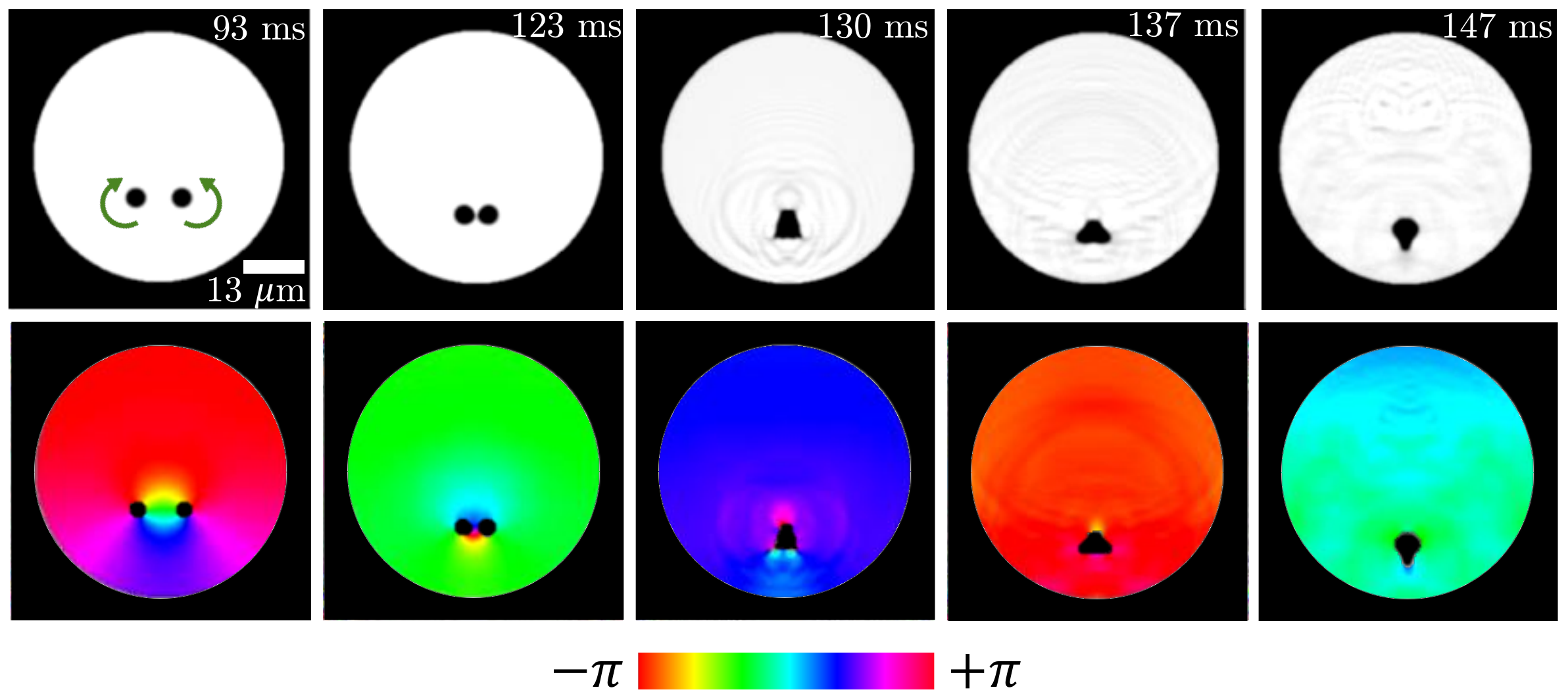}
    \caption{Mass-driven collision of a pair of counter-rotating massive vortices. Starting from a pair of singly-quantized vortices, upon colliding, the two vortices mutually annihilate and their incompressible kinetic energy is fully transferred into a sound pulse. First (second) row corresponds to the density (phase) field of the majority component ($^{87}{\rm Rb}$). Microscopic parameters for a mixture of $^{87}{\rm Rb}$ and $^{41}{\rm K}$ with $N_a=5\times 10^5,\,N_b= 5.4 \times 10^3,\, R=25\,\mu$m, $d_z=0.82\,\mu$m, $\omega_b = 85\, \mathrm{rad/s}$. The collision time $t_*$ is 125\,ms.}
    \label{fig:Counter_rotating_collisions}
\end{figure}

As a possible platform to implement our protocol we consider a heteronuclear mixture of $^{87}\mathrm{Rb}$ and $^{41}\mathrm{K}$ atoms, both in the internal state $|F,m_F\rangle=|1,+1\rangle$.
The two gases can be condensed in the same trap and present a favorable window of magnetic fields where the interspecies scattering length can be smoothly tuned and three-body losses minimized \cite{Burchianti2018}. In the following, Rubidium and Potassium represent component $a$ and component $b$, respectively, and in all the numerical analysis, we use $a_{aa}=99\, a_0$, $a_{bb}=65\, a_0$ and $a_{ab}=163\, a_0$, with $a_0$ the Bohr radius. For the strong confining harmonic potential along $z$, we use $d_z=0.82\,\mu$m.

We consider two massive vortices that, at time $t=0$, are symmetrically placed with respect to a diameter of the circular trap ($x_1=-x_2$, $y_1=y_2$) and are either at rest (opposite-charge case) or featuring a uniform circular precession \cite{Richaud2020} (same-charge case). In the massless case, the motion of these vortices is well-known and does not feature any collision: while the vortex dipole performs symmetric lobe-like orbits, each of them consisting of an almost straight segment and a circular arc \cite{Neely2010,Kwon2021}, the vortex pair exhibits a uniform circular orbit. These trajectories correspond to the light-colored lines in the left and right panels of Fig.~\ref{fig:Collision_sketch}. On the other hand, massive vortices subject to a species-selective confining potential, feel an additional force able to catalyzes the collision.  Solving the equations of motion~(\ref{eq:Motion_equation}), one can notice  (see dark-colored lines in Fig.~\ref{fig:Collision_sketch}) that the intervortex distance $r_{12}:=|{\bm r}_1 - {\bm r}_2|$ can tend to zero.
To our purposes, we stopped the solution of Eqs.~(\ref{eq:Motion_equation}) at the time $t_*$ such that $r_{12}=2\xi_a$, at which the two quantum vortices are expected to undergo an actual collisional event (for simplicity, we neglect the possible impact of component-$b$ particles on the vortex shape, i.e. on $\xi_a$).
Solving the full GPEs with the same initial condition of Eq.~(\ref{eq:Motion_equation}),  we find that the massive point vortex model gives very accurate results for $0<t<t_*$ (dots and solid lines in Fig.~\ref{fig:Collision_sketch} almost coincide). 

In Fig.~\ref{fig:Counter_rotating_collisions}, we illustrate the GPE simulated dynamics  of the density field $\rho_a$ (upper row) and the phase field $\theta_a$ (lower row) associated to the condensate wavefunction $\psi_a=\sqrt{n_a}e^{i\theta_a}$ for the case of two counter-rotating vortices (the wavefunction associated to the minority component, $\psi_b$, is not shown because it is simply non-zero only within the vortex cores).

\begin{figure}[t!]
    \centering
    \includegraphics[width=1\linewidth]{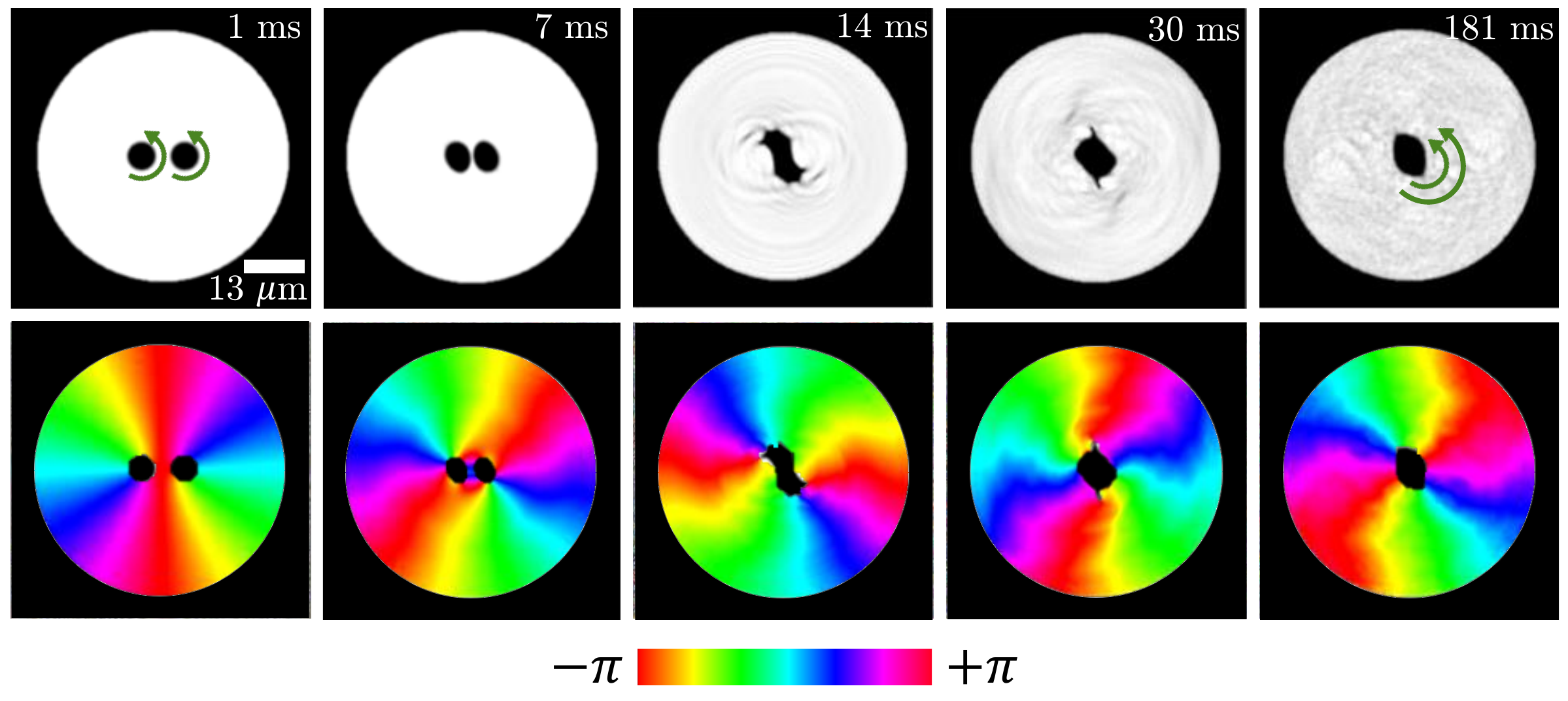}
    \caption{Mass-driven collision of a pair of co-rotating massive vortices. Starting from a pair of singly-quantized vortices, upon collision, a single doubly-quantized vortex is stabilized. Same microscopic model parameters [except that $N_b=1.35\times 10^4$ and $\omega_b=250 \, \mathrm{rad/s}$] and graphical conventions of Fig.~\ref{fig:Counter_rotating_collisions}. The collision time is $t_*=9$ ms.}
    \label{fig:Co_rotating_collisions}
\end{figure}

It is clear that, at $t=t_*$, the two quantum vortices collide and mutually annihilate. One can observe, in fact, that, for $t>t_*$, the density-depleted hole in $\rho_a$ is associated to no phase singularities (the corresponding phase field $\theta_a$ is indeed trivial) and, for this reason, it should be only ascribed to the persistent presence of $b$ particles which, being immiscible with respect to the $a$ fluid, induce a local spatial phase separation of the two components. On the other hand, a dramatic consequence of such annihilation process is the emission of an intense sound pulse, basically due to the conversion of vortex core and incompressible kinetic energies into phononic excitations \cite{Wlaszlowski_sound,Kwon2021}. This is particularly relevant in the context of quantum hydrodynamics \cite{Tsubota_Quantum_Hydrodynamics} and quantum turbulence \cite{Barenghi2014,Reeves2022}, as it constitutes a novel and unexplored mechanism supporting the annihilation of single vortex-antivortex pairs. 

The other fundamental class of two-vortex collisions involves two co-rotating vortices. In Fig.~\ref{fig:Co_rotating_collisions}, we illustrate the simulated GP dynamics, clearly showing the merging of two singly-quantized vortices into a doubly-quantized one. 
The latter state is known to be dynamically unstable \cite{Castin1999,Garcia1999,Butts1999} and prone to splitting in the massless case (see also Ref.~\cite{Kwon2021} for a real-time observation), but it is stabilized by the presence of component-$b$ particles. These particles, therefore, not only guide the collision of the two vortices as actively-driving pinning centers, but also stabilize the resulting doubly-quantized vortex. This result is far from being trivial and opens the door to the stabilization of multiply-quantized vortices in BECs, even in the absence of an external pinning potential \cite{Simula2002}. Unexpectedly, also in this case, the collision of the two quantum vortices is inelastic, as it comes with the emission of sound waves. This represents an additional mechanism supporting the conversion of incompressible kinetic energy into compressible kinetic energy.

\begin{figure}[t!]
    \centering
    \includegraphics[width=0.8\linewidth]{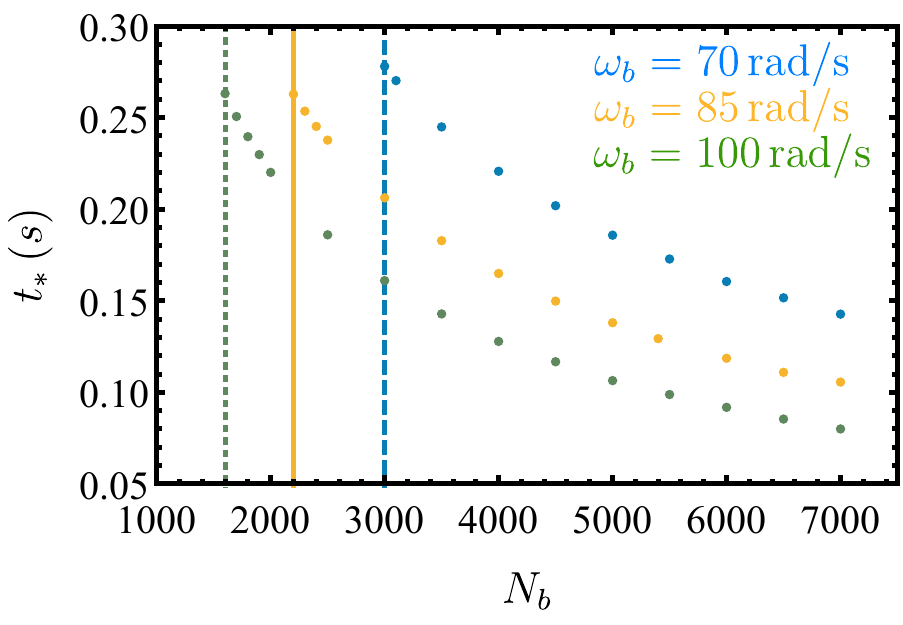}
    \caption{Increasing the number of core atoms ($N_b$), the vortices get heavier and the collisional time ($t_*$) decreases. Below a critical value of $N_b$ (vertical lines), cores are too light and vortices do not collide at all. The stronger the harmonic confinement acting on component-$b$ atoms, the lower are the collision time and the critical value for $N_b$. Same model parameters of Fig.~\ref{fig:Counter_rotating_collisions} were used.}
    \label{fig:Critical_time}
\end{figure}

To further substantiate the discussed collisional events and their being robust with respect to variations of microscopic parameters, we computed the collision time $t_*$ for different values of $N_b$ and $\omega_b$. As illustrated in Fig.~\ref{fig:Critical_time}, heavier core masses (larger values of $N_b$) correspond to  faster collisions, i.e., to smaller values of $t_*$. The latter can be further lowered by increasing $\omega_b$, the strength of the component-$b$ trapping potential, which we recall to be the catalyst of two-vortex collisions in the presented platform. Also, given a specific value of $\omega_b$, there exist a critical number $\tilde{N}_b$ of component-$b$ atoms below which the two vortices are too light to collide (see vertical lines in Fig.~\ref{fig:Critical_time}). In the Supplemental Material, we further comment on the relation $\tilde{N}_b(\omega_b)$, as well as on the robustness of massive vortices.

In the following, we propose a possible realistic experimental implementation for the observation of the merging and annihilation of massive vortices. A first requirement is immiscibility. 
This can be realized either using different species or different spin states. 

Depending on the kind of mixture, different methods and experimental techniques can be used. 
Besides being immiscible and stable (no spin relaxation or collisional decay channels), the mixture must allow for the possibility to handle it with component-selective potentials without excessive heating. 
As anticipated before, a mixture of $^{87}$Rb and $^{41}$K, both in the $|1,+1\rangle$ state, appears as a very promising candidate because the intraspecies scattering lengths are both positive and the interspecies one can be easily tuned around the Feshbach resonance at 78 G \cite{Burchianti2018} to select the proper immiscibility condition. Furthermore it is stable versus spin relaxations, both species being in a stretched spin state. The wavelengths for the $D_1$ and $D_2$ transitions for K and Rb are $\lambda^{\rm K}_{D_1,D_2}$=770 nm, 766.5 nm and $\lambda^{\rm Rb}_{D_1,D_2}$=795 nm, 780 nm. In this way, it is convenient to use Rb as component $a$ and selectively trap K with a focused Gaussian beam with a tune-in wavelength for Rb at 790 nm  \cite{Catani2009,Lamporesi2010,Catani2012}, both for the initial positioning of the minority $b$ component and also for the species-selective harmonic confinement. The mixture can be initially trapped in a flat-box optical trap, realizable with DMD technology \cite{Navon2021} and able to confine both components in a circular region. A repulsive optical barrier can be spun \cite{Write2013} around such large impurities clockwise or counterclockwise depending on the vortex charge which one wants to have in the majority component $a$. Alternatively, vorticity can be introduced through phase imprinting, by illuminating the atomic sample with a light pattern corresponding to the desired phase pattern \cite{delPace2022}. In this case the regions where phase $0$ and $2\pi$ connect should be depleted using a repulsive potential to avoid unwanted smooth phase profiles in the wrong direction \cite{Kumar2018}. Once vortices are created in component $a$, the pinning potential for component $b$ can be removed while activating the harmonic potential to study the vortex dynamics. 
In the alternative case of using a spin mixture of a single atomic species, with a proper choice of magnetic states, one could also use a magnetic harmonic trap concentric with the optical box and prepare
species $a$ in a magnetic insensitive state $m_F=0$, while $b$ in a low-field-seeker state. 
The density of the minority component can be directly imaged in the plane to observe the collision. Furthermore one can measure the vorticity in the majority component after the merging event, through Bragg interferometric measurements \cite{Donadello2014} or by removing the species selective potential and observing the formation of singly quantized vortices.

In conclusion, we have shown that the inertial effect of particles deliberately or accidentally trapped in the cores of quantum vortices opens the door to two-vortex collisions, a scenario that has no counterpart within standard superfluid vortex dynamics \cite{Neely2010,Seo2017}. We have presented a realistic experimental platform involving a quasi-2D imbalanced mixture of BECs which allows one to design and observe two-vortex collisions in a controlled fashion. The proposed protocol relies on the use of a component-selective potential \cite{Catani2009,Lamporesi2010,Catani2012}, which catalyzes the occurrence of collisions in two-vortex systems. The massive cores do not simply alter standard vortices trajectories \cite{Richaud2021}, but play the role of effective pinning potentials driving quantum vortices to collide. We demonstrate the mutual annihilation of vortex dipoles and the ensuing radiation of sound pulses, as well as the merging of two singly-quantized vortices resulting in the stabilization of a doubly-quantized vortex in the absence of an external pinning \cite{Simula2002}. 
In addition, our results provide a clear explanation to previous numerical observations of half-quantum vortex collisions \cite{Kasamatsu2016}. This interpretation will also be important to get a better insight about the superfluid turbulent behavior of such a system \cite{Wheeler2021}.

The studied phenomenology is expected to be rather general, as it originates only from the presence of trapped particles within -- or equivalently a mass of -- the quantum vortices, a circumstance which is quite common in many superfluid and magnetic systems.

\begin{acknowledgements}
We thank C. Fort for providing useful data on the K-Rb mixture and P. Massignan for the stimulating discussions.
We acknowledge funding by Provincia Autonoma di Trento and from MIUR through the PRIN 2017 (Prot. 20172H2SC4 005)  and PRIN 2020 (Prot. 2020JLZ52N 002) programs.
\end{acknowledgements}

\clearpage

\widetext
\begin{center}
\textbf{\large Supplemental Materials: Mass-driven vortex collisions in flat superfluids}
\end{center}

\twocolumngrid

\setcounter{equation}{0}
\setcounter{figure}{0}
\setcounter{table}{0}
\setcounter{page}{1}
\makeatletter
\renewcommand{\theequation}{S\arabic{equation}}
\renewcommand{\thefigure}{S\arabic{figure}}
\renewcommand{\bibnumfmt}[1]{[S#1]}
\renewcommand{\citenumfont}[1]{S#1}

\section*{The massive point vortex model}
In the Thomas-Fermi (TF) regime, vortices can be regarded as point-like defects, a circumstance which implies that the overall density $n_a(x,y)$ is essentially uniform within the considered hard-wall trap of radius $R$, except in a small region of radius $\sim \xi_a$ centered at the vortex position ${\bm r}_j=(x_j,\,y_j)$. On the other hand, the presence of a vortex of charge $q_j$ in the atomic cloud has a profound impact on the  phase field $\theta_a(x,y)$ of the condensate (we recall that $\psi_a=\sqrt{n_a}e^{i\theta_a}$). It is well known that, in the case of a circular potential well with hard walls, the velocity field ${\bm v}=\frac{\hbar}{m_a}\nabla \theta_a$ associated to the condensate-$a$ wavefunction is such that its normal component must vanish at the boundary of the trap \cite{Kim2004}. This condition is ensured by the presence of image vortices of opposite circulation at the external positions ${\bm r}_j^\prime={\bm r}_j R^2/r_j^2$.

A standard way to capture the dynamics of the field $\psi_a$ in the presence of $N_v$ vortical excitations is to develop a time-dependent variational approach \cite{Garcia1996,Kim2004,Richaud2020}. One makes a time-dependent variational Ansatz for the field $\psi_a=\sqrt{n_a}e^{i\theta_a}$, which, in our case, is given by 
\begin{equation}
\label{eq:Ansatz_theta_a}
    \theta_a(x,y)=\sum_{j=1}^{N_v} q_j\left[\arctan\left(\frac{y-y_j}{x-x_j}\right)-\arctan\left(\frac{y-y_j^\prime}{x-x_j^\prime}\right)\right].
\end{equation}
and where the vortices coordinates $\bm{r}_j(t)=(x_j(t),\,y_j(t))$ are the only parameters assumed to depend on time. Then, one computes the Lagrangian functional 
\begin{equation}
\label{eq:L_psi_a}
   \mathcal{L}_a[\psi_a]=\mathcal{T}_a[\psi_a]-\Delta\mathcal{E}_a[\psi]
\end{equation}
where 
\begin{equation}
\label{eq:T_k_psi}
    \mathcal{T}_a[\psi_a]=\int \frac{i\hbar}{2}\left(\psi_a^* \frac{\partial\psi_a}{\partial t}-\frac{\partial \psi_a^*}{\partial t}\psi_a\right)\, \mathrm{d}^2r
\end{equation}
is the kinetic term and
\begin{equation}
\label{eq:E_k_psi}
   \Delta\mathcal{E}_a[\psi_a]=\int |\psi_a|^2 \frac{1}{2} m_av_a^2 \, \mathrm{d}^2r=\int n_a \frac{\hbar^2}{2 m_a} (\nabla \theta_a)^2 \, \mathrm{d}^2r,
\end{equation}
is the energy difference with respect to the vortex-free state. 
$\Delta\mathcal{E}_a$ is formally divergent in a vorticous flow, because the superfluid velocity behaves as $1/|\bm{r}-\bm{r}_j|$ around each vortex core. To regularize it, one customarily excludes a circular region of radius $\xi_a$ around each vortex core. 

One thus remains with an effective point-like Lagrangian which governs the dynamics of the assumed time-dependent variational parameters and thus provides an effective description of the system's dynamics. The application of the time-dependent variational approach to a many-vortex system in a quasi 2D Bose-Einstein condensate is described in Ref.~\cite{Kim2004}.

The presence of massive particles trapped within the vortices core modifies the ``bare" vortex dynamics both because of their inherent inertial contribution \cite{Richaud2020,Richaud2021} and because of the (radial) force $F=-\partial V_{\rm tr}^b /\partial r=- m_b\omega_b^2 r$ which component-$b$ particles are subject to. Both effects can be incorporated into the (massless) point vortex model (\ref{eq:L_psi_a}). As in Refs. \cite{Garcia1996,Richaud2021}, for the localized $b$-component, we introduce the following time-dependent Gaussian ansatz:  
\begin{equation}
\label{eq:Gaussian_ansatz}
\psi_b(\bm r) =\sum_{j=1}^{N_v}  \sqrt{\frac{N_b}{\pi\sigma^2 N_v}}e^{-|\bm r - \bm r_j(t)|^2/2\sigma^2}e^{i\bm r\cdot \bm \alpha_j(t)}
\end{equation}
representing an array of $N_v$ localized wave packets with width $\sigma$ centered at $\bm r_j(t)$ and where the parameter $\bm\alpha_j(t)$ allows for a nonzero translational velocity $\dot{\bm r}_j = \hbar\bm \alpha_j/m_b$. In principle, the $b$-component core size $\sigma$ depends on a number of microscopic parameters, including interaction constants ($g_a$, $g_b$, $g_{ab}$), the number of atoms in the two components ($N_a$, $N_b$), their atomic masses ($m_a$, $m_b$), the trap radius ($R$) and the condensate thickness ($d_z$) \cite{Richaud2020,Braz2020}, but, for the point model to be valid, one only requires that $\sigma \ll R$.

A straightforward analysis based on the time-dependent variational approach reviewed above (see also Refs.~\cite{Garcia1996,Richaud2021}) leads to the effective point-like Lagrangian for species-$b$ cores
\begin{equation}
\label{eq:L_b_final}
\mathcal{L}_b =  \sum_{j=1}^{N_v}\left[\frac{N_b m_b}{2 N_v}\dot{\bm r}_j^2 -\frac{1}{2} \frac{N_b m_b}{N_v} \omega_b^2 r_j^2\right],
\end{equation}
which depends on both the average position $\bm r_j$ and average velocity $\dot{\bm r}_j \propto \bm \alpha_j$ of each wave packet in Eq. (\ref{eq:Gaussian_ansatz}). Notice that this Lagrangian includes both the Newtonian inertial term $\propto \dot{\bm r}_j^2$ and the effect of the external confining potential $\propto r_j^2$.  

As explained in Ref.~\cite{Richaud2021}, the effective dynamics of the overall composite system, i.e., of component-$a$ vortices filled by component-$b$ particles, can be described in terms of the following Lagrangian, here specialized to the case of $N_v=2$ vortices
$$
    \label{eq:L_tot}
    \mathcal{L}_{\rm tot} = \mathcal{L}_a + \mathcal{L}_b
$$
\begin{equation}
= \sum_{j=1}^2 \left[ \frac{M_j}{2}\dot{\bm r}_j^2 + \frac{k_j\rho_a}{2} \dot{\bm r}_j \times {\bm r}_j \cdot \hat{z}  \right] - V(\bm{r}_1,\bm{r}_2),
\end{equation}
where
$$
  V=\frac{\rho_a}{4\pi}\left\{k_1k_2 \ln  \left(\frac{R^2 - 2\bm r_1\cdot\bm r_2  + r_1^2r_2^2/R^2}{r_1^2 -  2\bm r_1\cdot\bm r_2  + r_2^2}\right)
   \right.
$$
\begin{equation}
   \left. +k_1^2\ln\left(1-\frac{r_1^2}{R^2}\right)  +k_2^2\ln\left(1-\frac{r_2^2}{R^2}\right) \right\}  + \sum_{j=1}^2 \frac{1}{2}M_j \omega_b^2 r_j^2.
\end{equation}

\section*{Robustness of massive vortices}
According to the collisional protocol proposed in the main text, two vortices are led to collide by their massive cores, which thus play the role of actively-driving pinning potentials. As a result, the trajectory ${\bm r}_j(t)$ of these vortices significantly differ from the one predicted by standard superfluid dynamics (compare the dark and light lines in Fig.~\ref{fig:Collision_sketch}). This implies that the massive core applies a non-negligible force on its hosting quantum vortex (and vice versa). If this force exceeds a certain critical value, the quantum-vortex--massive-core complex breaks down. More specifically, the massive core penetrates the vortex walls and enters the bulk of component-$a$ condensate. The depinning of a massive core from its quantum vortex is a complex phenomenon, and its description requires a detailed analysis of the surface tension of component-$a$ vortex walls, the compressibility properties of the majority component, as well as the miscibility of the two quantum fluids.

\begin{figure}
    \centering
    \includegraphics[width=0.75\linewidth]{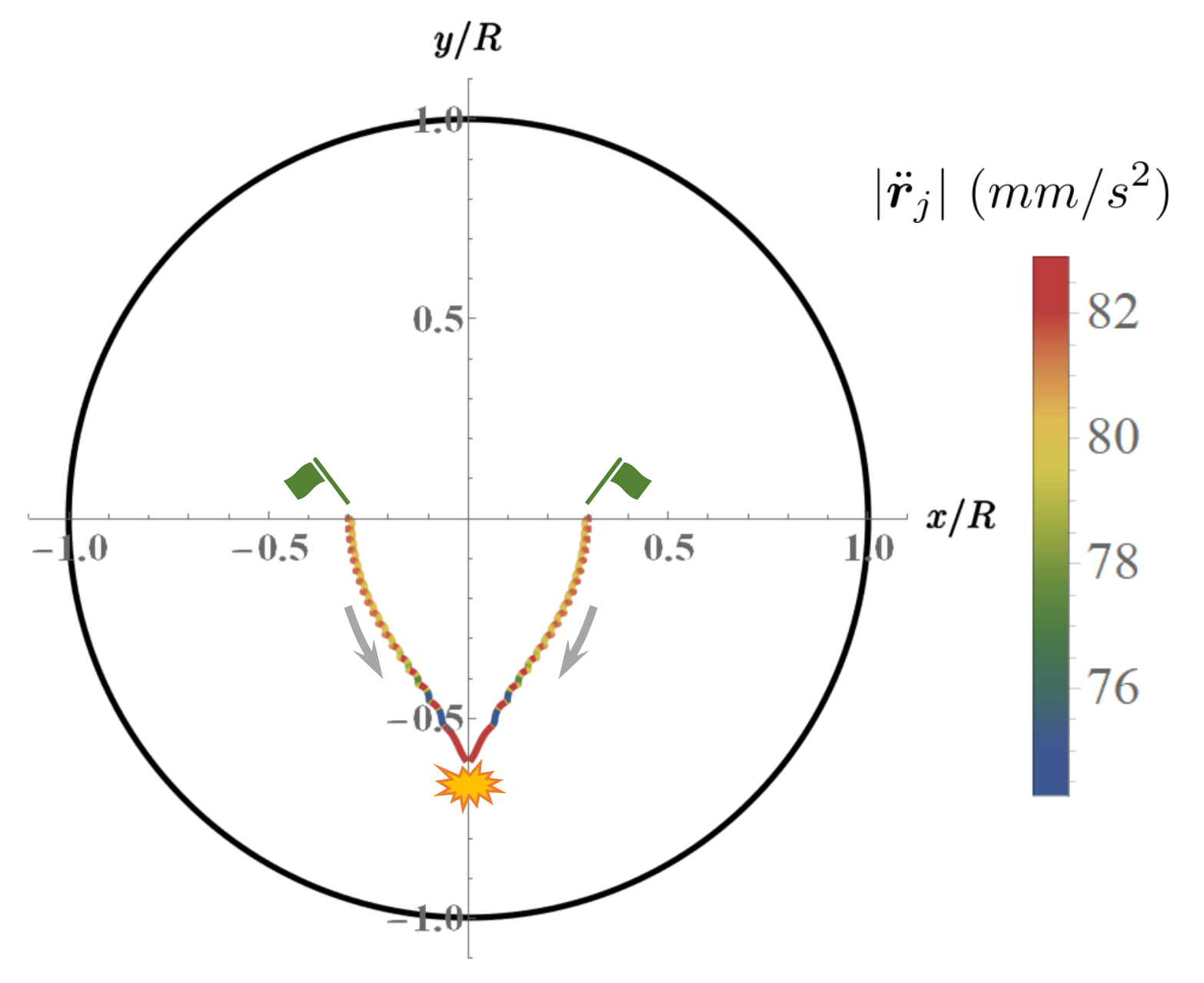}
    \caption{The (magnitude of) the acceleration $|\ddot{\bm{r}}_j|$ of the two massive vortices along their collisional trajectories is maximum at the start (green flags), just before the collision (orange star) and at the changes of direction. Model parameters coincide with the ones used for Figs.~\ref{fig:Collision_sketch},~\ref{fig:Counter_rotating_collisions}, and~\ref{fig:Co_rotating_collisions}.}
    \label{fig:Acceleration}
\end{figure}

Although we are going to fully characterize this physics in a separate work, we anticipate that the discussed depinning occurs in regimes different from the ones presented in the main text. In essence, the rather-large (im)miscibility ratio $g_{ab}/\sqrt{g_a g_b}\approx 2.18$ characterizing our $^{87}\mathrm{Rb}$-$^{41}\mathrm{K}$ mixture, together with the rather small values of the applied $\omega_b$, ensure the robustness of our massive vortices, as well as the smooth running of the designed collisional protocol. We remark, in this regard, that smaller values of $g_{ab}$ would lead to more fragile quantum-vortex--massive-core complexes, i.e., to structures which would be more prone to undergo depinning when subject to an acceleration. The acceleration $|\ddot{\bm r}_j (t)|$ experienced by massive vortices along their trajectories, from the very beginning of the collisional protocol to the collision itself, can be easily computed by means of Eq.~(\ref{eq:Motion_equation}). In Fig.~\ref{fig:Acceleration}, we illustrate the magnitude of this acceleration for the case of two counter-rotating massive vortices (we employ the same model parameters used for the left panel of Fig.~\ref{fig:Collision_sketch} and for Fig.~\ref{fig:Counter_rotating_collisions}).

\begin{figure}[b!]
    \centering
    \includegraphics[width=0.95\linewidth]{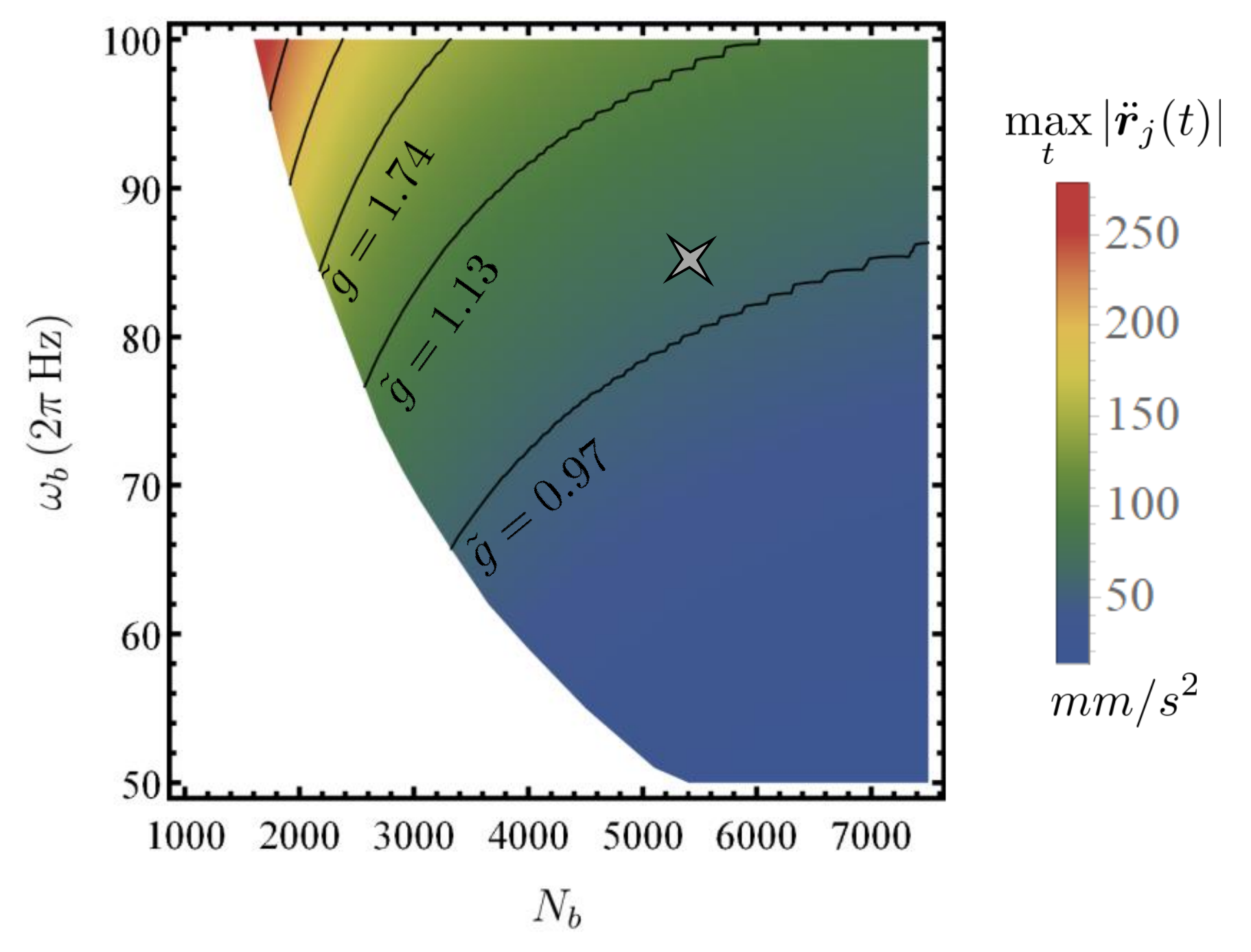}
    \caption{Maximum acceleration experienced by a massive vortex before colliding $(\max_t|\ddot{\bm r}_j(t)|)$ according to the collisional protocol illustrated in the left panel of Fig.~\ref{fig:Collision_sketch} and Fig.~\ref{fig:Counter_rotating_collisions}. The same model parameters are assumed. The white region correspond to the absence of collisional events and its boundary represents the relation $\tilde{N}_b(\omega_b)$ mentioned in the main text.}
    \label{fig:Phase_diagram}
\end{figure}

The red color denotes those portions of trajectories where massive vortices are subject to the strongest stress, i.e., where the force exerted on the vortex walls by the massive core is maximum. Such portions correspond to the start of the dynamics, to the changes of direction (cusp-like features), and to the collision itself (orange star). They represent the segments where the depinning of a massive core from its quantum vortex may occur and their extent depends on the value of the component-$b$ trapping frequency, $\omega_b$. It is worth remarking, in this regard, that, if on the one hand, higher values of $\omega_b$ correspond to smaller collision times ($t_*$) (see Fig.~\ref{fig:Critical_time}), on the other hand, massive vortices would experience larger accelerations which could, in principle, break them before they come to collide. In Fig.~\ref{fig:Phase_diagram}, we plot the maximum acceleration experienced by a massive vortex before colliding,  $\max_t\{|\ddot{\bm r}_j(t)|\}$.

Interestingly enough, each isoline can be associated to a critical value of $\tilde{g}:=g_{ab}/\sqrt{g_ag_b}$ below which the massive core would undergo depinning before colliding. As visible in the figure, for the set of model parameters which we considered (gray cross), and given that the immiscibility ratio of our mixture is 2.18, our quantum-vortex--massive-core complexes are well within the stability region and hence can be safely subject to the accelerations needed to carry out the proposed collisional protocol without the risk of breaking.

\end{document}